\documentclass[twocolumn,showpacs,preprintnumbers,
amsmath,amssymb,floatfix,nofootinbib]{revtex4-2}

\usepackage{dcolumn}
\usepackage{bm}
\usepackage{amsthm}
\usepackage{amsfonts}
\usepackage{anysize}

\usepackage{amsmath}
\usepackage{mathtools}
\usepackage[english]{babel}
\usepackage{graphicx}

\usepackage{caption}
\usepackage{subcaption}

\usepackage{amssymb}

\newcommand{\be}{\begin{equation}}
\newcommand{\ee}{\end{equation}}
\newcommand{\ba}{\begin{eqnarray}}
\newcommand{\ea}{\end{eqnarray}}
\newcommand{\non}{\nonumber}

\newcommand{\ce}{{\cal{E}}}
\newcommand{\cl}{{\cal{L}}}

\newcommand{\nin}{\noindent}

\usepackage[dvipsnames]{xcolor}

\begin{document}

\title{Circular orbits of Charged Particles around a Weakly Charged and Magnetized Schwarzschild Black Hole}
\author{A. M. Al Zahrani}
\email{amz@kfupm.edu.sa}
\email{ama3@ualberta.ca}
\affiliation{Physics Department, King Fahd University of Petroleum and Minerals, Dhahran 31261, Saudi Arabia}

\begin{abstract}
We study the circular orbits of charged particles around a weakly charged Schwarzschild black hole immersed in a weak, axisymmetric magnetic field. We start by reviewing the circular orbits of neutral particles and charged particles around only weakly charged and only weakly magnetized black holes. The case of a weakly magnetized and charged black hole is investigated then. In particular, we study the effect of the electromagnetic forces on the charged particles innermost stable circular orbits. We show that negative energy stable circular orbits are possible and that two bands of charged particles circular orbits, separated by a gap of no stable circular orbits can exist. The astrophysical aspects of our findings are discussed too.
\end{abstract}


\maketitle

\section{Introduction}

Black hole astrophysics has been one of the most intriguing disciplines of science. It reached the pinnacle of fascination when the first real image of a black hole was published by the Event Horizon Telescope Collaboration. One of the principal objectives of black hole astrophysics is to measure the parameters of a black hole, in particular, its mass and spin angular momentum. According to the no-hair-theorem, a stationary black hole is characterized by only three parameters, mass, spin angular momentum and charge~\cite{MTW}. It has been argued that astrophysical black holes are indeed electrically neutral. If some excess charge is produced by whatever astrophysical processes, it would be shortly neutralized be the selective accretion of the ambient plasma. The wide adoption of this assumption can be explicitly seen in the literature where charged black holes are hardly considered, except for academic purposes.

There are several profound reasons to assume that weakly charged black holes exist, however. The mass difference between electrons and protons may render a black hole positively charged (see Refs.\cite{Zaj1,Zaj2,Zaj3} and the references within). Moreover, a rotating black hole immersed into a homogeneous magnetic field acquires a non-vanishing electrical field. This can result in the selective accretion of the ambient free charged particles, until the black hole's charge neutralize the magnetically-induced electric field \cite{Carter, Wald}. Another effect that might charge a black hole is the difference in the accretion rates of electrons and protons of the surrounding plasma in the presence of radiation from the accreting matter.

It is widely accepted that astrophysical black holes are magnetized. The main source of the magnetic fields is the plasma in the accretion disk as discussed in Refs. \cite{RoVi,Pun}. In addition to the theoretical considerations, there is abundant observational data that backs this assertion \cite{Ruth,sil1,sil2,sil3,sil4}. 

The innermost stable circular orbit (ISCO) of a black hole has several immediate implications. It is vital for measuring the spin angular momentum of the black hole \cite{Bren}. It is also influential to the structure of the accretion disk \cite{AF}. For an accretion disk with low luminosity compared to the Eddington luminosity, the ISCO coincides with the inner edge of the disk \cite{Ab}. Therefore, the ISCO has a direct impact on the structure of the black hole's shadow.

The charged particles ISCOs around a Schwarzschild and Kerr black holes immersed in a weak, axially symmetric magnetic field were extensively investigated \cite{GP,AG,AO,Z}. In all cases, the effect of the magnetic field is to bring the ISCO closer to the black hole. The dynamics of charged particles in Reissner-Nordström spacetime was investigated in \cite{PQR,PQR2}. The charged particles ISCOs around a weakly charged Schwarzschild black hole were addressed in \cite{Zaj1,Zaj2,Zaj3}. The combined affects of the magnetic field and black hole's charge on the ISCOs have been recently examined in Ref.~\cite{HH} for a few special cases.

In this paper we study the ISCOs of charged particles orbiting a Schwarzschild black hole that is weakly charged and immersed in a weak, axisymmetric magnetic field. The black hole's charge and magnetic field are weak in the sense that their back-reactions on the spacetime are insignificant\footnote{In what follows, it should be inferred that the black hole's charge and magnetic field are assumed to be weak.}. We first look at the effect of each of the magnetic field and black hole's charge separately, and then study their combined effect. We explore in detail the cases when no ISCO exists and when the particle's energy become negative. The paper is organized as follows: Sec.~\ref{s2} is a review of the circular orbits and the ISCO of a neutral particle. In Sec.~\ref{s3}, we go over the Wald's solution of Maxwell's equations in Ricci flat spacetimes and write the radial equation of motion for a charged test particle. In Secs.~\ref{s4} and \ref{s5} we discuss the ISCOs of a charged particle near a charged black hole and magnetized black hole, respectively. In Sec.~\ref{s6}, we investigate in detail the charged particle ISCOs near a charged black hole immersed in a magnetic field. General discussion and conclusion is given in Sec.~\ref{sum}. We use the sign conventions adopted in Ref.~\cite{MTW} and geometrized units where $c$, $G$ and $k$ (the Coulomb constant) are unity.

\section{Circular Orbits of a Neutral Particle around a Schwarzschild Black Hole} \label{s2}

The spacetime geometry around a spherically symmetric black hole of mass $M$ is described by the Schwarzschild metric, which reads~\cite{MTW}
\begin{eqnarray}
ds^2=&-&f(r)dt^2+\frac{dr^2}{f(r)}+r^2 d\theta^2 \non \\
     &+&r^2\sin^2\theta d\phi^2 \:\:,
\end{eqnarray}
where
\begin{eqnarray}
f(r) = 1-\frac{2M}{r}
\end{eqnarray}

\nin The Schwarzschild metric admits four Killing vectors. Two of them are the temporal and azimuthal Killing vectors, since the metric is temporally and azimuthally symmetric. They respectively read
\begin{equation}
\xi^{\mu}_{(t)}=\delta^{\mu}_t, \:\:\: \xi^{\mu}_{(\phi)}=\delta^{\mu}_{\phi}.
\end{equation}

\nin Let a test particle of mass $m$ be moving with four-velocity $u^{\mu}$ in the Schwarzschild background. There are two constants of the particle's motion associated with the two Killing symmetries:
\begin{eqnarray}
\ce &=& -p_{\mu}\xi^{\mu}_{(t)}/m=f(r)\dot{t}, \label{en0} \\
\cl &=& p_{\mu}\xi^{\mu}_{(\phi)}/m=r^2\sin^2{\theta}\dot{\phi}.\label{am0}
\end{eqnarray}

\nin Here $p^{\mu}=mu^{\mu}$ is the particle's four-momentum. The two constants of motion $\ce$ and $\cl$ are the specific energy and specific azimuthal angular momentum, respectively. Using them along with the normalization $u_{\mu}u^{\mu}=-1$, we reduce the radial equation of motion in the equatorial submanifold ($\theta=\pi/2$, $\dot{\theta}=0$) to quadrature:
\begin{equation}
\dot{r}^2=\ce^2-V(r) \label{rdot0}.
\end{equation}
\nin The overdot denotes differentiation with respect to the particle's proper time. The effective potential $V(r)$ reads
\begin{eqnarray}
V(r) = \left(1-\frac{\cl^2}{r^2}\right)f(r) \label{pot0}.
\end{eqnarray}

\nin The radial motion is  invariant under the transformations
\begin{equation}\label{sym}
\phi\rightarrow-\phi, \hspace{3mm} \dot{\phi}\rightarrow-\dot{\phi}, \hspace{3mm} \cl\rightarrow-\cl.
\end{equation}

\nin Therefore, there is only one mode of radial motion. Without loss of generality, we will consider $\cl>0$.

\nin The two conditions of circular motion $V(r)=\ce$ and $V'(r)=0$ give us
\begin{eqnarray}
&\ce^2&=\left(1-\frac{\cl^2}{r^2}\right)f(r),\\
&\ce^2&(r-M)+(2r-3M)r^2=0.
\end{eqnarray}

\nin We will use $r_o$, $\ce_o$ and $\cl_o$ to denote quantities corresponding to circular orbits from here on. Solving the above two equations for $\ce_o$ and $\cl_o$ for future-directed orbits yields
\begin{eqnarray}
\ce_o&=&\frac{(r_o-2M)}{r_o^{1/2}\sqrt{r_o-3M}},\label{ec}\\
\cl_o&=&\frac{Mr_o}{\sqrt{r_o-3M}}.\label{lc}
\end{eqnarray}

\nin The value of $\ce_o$ is always positive with a value of $1$ far away from the black hole. A circular orbit is the ISCO when $V''(r_o)$ vanishes. This condition gives us that $r_{\text{ISCO}}=6M$. It is worth mentioning that $\ce_{\text{ISCO}}=2\sqrt{2}/3$ and $\cl_{\text{ISCO}}=2\sqrt{3}M$. Therefore, a particle ending in the ISCO can release an energy of $1-2\sqrt{2}/3\: (\approx 0.057)$ of its rest energy.

\section{Charged and Magnetized Schwarzschild Black Holes}\label{s3}

Let us now review Wald's solution of Maxwell equations in a curved space time for weak electromagnetic fields \cite{Wald}. In a Ricci flat spacetime a Killing vector $\xi^{\mu}$ obeys the equation
\begin{equation}
\xi^{\mu\:\:\:\: ;\nu}_{\:\:\;;\nu}=0.
\end{equation}

\noindent This is identical to the source-free Maxwell equations for a four-potential $A^{\mu}$ in the Lorentz gauge ($A^{\mu}_{\:\:\:;\mu}=0$),
\begin{equation}
A^{\mu\:\:\:\: ;\nu}_{\:\:\;;\nu}=0.
\end{equation}
Therefore, any linear combination of the Killing vectors the spacetime admits is automatically a solution to the Maxwell equations.

\nin Consider the electromagnetic potential constructed of the temporal and azimuthal Killing vectors of the Schwarzschild geometry:
\begin{equation}
A^{\mu}=-\frac{Q}{2M}\xi^{\mu}_{(t)}+\frac{B}{2}\xi^{\mu}_{(\phi)}.
\end{equation}
\nin Lowering the index and removing the constant term give
\begin{equation}\label{empot}
A_{\mu}=-\frac{Q}{r}\:\delta_{\mu}^t+\frac{B}{2}r^2\sin^2\theta\:\delta_{\mu}^{\phi}.
\end{equation}

\noindent This potential describes the electromagnetic fields around a charged black hole immersed in an axisymmetric magnetic field. The black holes's charge $Q$ is given by 
\begin{equation}\label{empot}
Q=\frac{1}{4\pi}\int_{\sigma}F^{\mu\nu}d\sigma_{\mu\nu},
\end{equation}
\nin where $\sigma$ is a 2D surface surrounding the black hole and $F^{\mu\nu}$ is the electromagnetic field tensor (see Eq.~\ref{emt}). The magnetic field is axisymmetric with a strength of $B$ asymptotically \cite{Wald,AG,AO}. This is the potential that we will use in this paper. 

The dynamics of a charged particle of mass $m$ and charge $e$ in an electromagnetic field in a curved spacetime is governed by the equation
\begin{equation}\label{de}
m{u}^{\nu}\nabla_{\nu}u^{\mu}=eF^{\mu}_{\;\;\rho}u^{\rho}.
\end{equation}

\noindent The electromagnetic field tensor $F_{\;\;\nu}^{\mu}$ is given by
\begin{equation}\label{emt}
 F_{\mu\nu}=A_{\nu,\mu}-A_{\mu,\nu}.
\end{equation}

\nin In the frame of an observer with four-velocity $u_\mu^{\text{obs}}$, the electric and magnetic fields are, respectively
\begin{eqnarray}
E^{i}&=&F^{i\nu}u_{\nu}^{\text{obs}}, \\
B^{\mu}&=&-\frac{1}{2}\frac{\varepsilon^{\mu\nu\lambda\sigma}}{\sqrt{-g}}F_{\lambda\sigma}u_{\nu}^{\text{obs}},
\end{eqnarray}
where $g=\mbox{det}(g_{\mu\nu})$, $\varepsilon_{0123}=+1$ and $i = 1,2,3$.
\nin For a stationary observer ($ u_{\mu}^{obs}=-f^{1/2}\delta^t_{\mu}$), the electric and magnetic fields are
\begin{eqnarray}
E^{i}&=&\frac{Qf^{1/2}}{r^2}\delta_r^i, \\
B^{i}&=&Bf^{1/2}\left(\cos{\theta} \: \delta^i_r-\frac{\sin\theta}{r}\: \delta^i_{\theta}\right),
\end{eqnarray}

\nin The generalized four-momentum of the particle is
\begin{equation}
P_{\mu}=mu_{\mu}+eA_{\mu}.
\end{equation}

\nin The Lie derivatives of $A_{\mu}$ with respect to $\xi_{(t)}^{\mu}$ and $\xi_{(\phi)}^{\mu}$ identically vanish:
\begin{eqnarray}
{\cal L}_{\xi^{\nu}_{(t)}}A^{\mu}&=&0, \\
{\cal L}_{\xi^{\nu}_{(\phi)}}A^{\mu}&=&0.
\end{eqnarray}
\nin The energy and azimuthal angular momentum of a charged particle are therefore constants of motion. Eqs.~\ref{en0} and~\ref{am0} are generalized to 
\begin{eqnarray}
{\cal E}&=&-P_{\mu}\xi^{\mu}_{(t)}/m = \frac{q}{r}+f(r)\dot{t},  \label{en1}\\
{\cal L}&=&P_{\mu}\xi^{\mu}_{(\phi)}/m =r^2(b+\dot{\phi})\sin^2{\theta}. \label{am1}
\end{eqnarray} \\
\nin where $q={eQ}/{m}$ and $b={eB}/{2m}$. We can straightforwardly obtain the charged particle version of Eq.~\ref{rdot0} by combining Eqs.~\ref{en1} and~\ref{am1} with $u_{\mu}u^{\mu}=-1$. The radial equation of motion in the equatorial submanifold then reads
\begin{equation}
r^2\dot{r}^2=\left(\ce r-q\right)^2-\left[r^2+\left(\cl-br^2\right)^2\right]f(r).\:\:\: \label{rdot}
\end{equation}
\nin It is more convenient to recast it as
\begin{equation}
\dot{r}^2=\left(\ce-V_+\right)\left(\ce-V_-\right),
\end{equation}
\nin where
\begin{equation}
V_{\pm}(r) = \frac{q}{r}\pm\sqrt{\left[1+\left(\frac{\cl}{r}-br\right)^2\right]f(r)}. \label{pot}
\end{equation}
It is $V_+(r)$ that corresponds to future-directed orbits. Eq.~\ref{rdot} is invariant under the symmetry transformations
\begin{equation}
\phi\rightarrow-\phi,\;\;\; \dot{\phi}\rightarrow-\dot{\phi},\;\;\; \cl\rightarrow-\cl,\;\;\; b\rightarrow -b. \label{sym2}
\end{equation}
As in the previous section, we will keep $\cl>0$ without any loss of generality. When $b>0$ ($b<0$), the magnetic force is radially out (in). Likewise, $q>0$ ($q<0$) corresponds to Coulomb repulsion (attraction). Therefore, there four different modes of radial motion, in general. 

The weak field approximation breaks down when the electric charge and magnetic field creates curvatures comparable to that made by the black hole's mass near the even horizon. This happens when
\begin{equation}
B^2\sim M^{-2} \text{ or } Q^2\sim M^2.
\end{equation}

\nin In conventional units, the weak field approximation fails when
\begin{equation}
Q\sim \frac{G^{1/2}M}{k^{1/2}} \sim 10^{20} \frac{M}{M_\odot}\; \text{coulomb},
\end{equation}
\nin or
\begin{equation}
B\sim \frac{k^{1/2}c^3}{G^{3/2}M} \sim 10^{19} \frac{M_\odot}{M}\; \text{gauss},
\end{equation}
where $M_\odot$ is the solar mass.
\nin The typical magnetic field strength near a black hole's horizon has been estimated to be $\sim10^8$ gauss ($10^{-15} \:\text{meter}^{-1}$) for stellar mass black holes and $\sim10^4$ gauss ($10^{-19} \:\text{meter}^{-1}$) for supermassive black holes~\cite{Ruth,sil1,sil2,sil3,sil4}. According to Ref.~\cite{Zaj2}, the charge of Sgr~A* is estimated to be in the range $10^8$--$10^{15}$ ($10^{-9}$--$10^{-2}$ meter). These estimates validate ignoring corrections to the metric due to the presence of the electromagnetic fields. 

\nin In spite of the fact that the electromagnetic fields are geometrically insignificant, their effects on the dynamics of charged particles can be significant since $e/m=2.04\times 10^{21}\; (1.11\times 10^{18})$ for electrons (protons). For electrons and protons near a black hole with $Q=10^8\; \text{coulomb}$ and $B=10^4\; \text{gauss}$, for example, 
\begin{equation}
q_{\text{e}}\sim 10^{12}\;\text{meter}, \;\;\; q_{\text{p}}\sim 10^{9}\; \text{meter},
\end{equation}
and
\begin{equation}
b_{\text{e}}\sim 10^{3}\;\text{meter}^{-1}, \;\;\; b_{\text{p}}\sim 10^{-1}\; \text{meter}^{-1}.
\end{equation}
The subscripts "e" and "p" refer to electrons and protons, respectively.

\section{Circular Orbits around a Charged Schwarzschild Black Hole}\label{s4}

\nin Now, we write the expressions similar to Eqs.~\ref{ec} and~\ref{lc} for a charged particle orbiting a charged Schwarzschild black hole and study its circular orbits and ISCOs. Setting $b=0$ in Eq.~\ref{pot} gives
\begin{equation}
V_+(r) = \frac{q}{r}+\sqrt{\left(1+\frac{\cl^2}{r^2}\right)\left(1-\frac{2M}{r}\right)}.
\end{equation}
The two conditions for circular orbits  ($V(r)=\ce$,  $V'(r)=0$) respectively yield
\begin{equation}
\ce_o = \frac{q}{r_o}+\sqrt{\left(1+\frac{\cl_o^2}{r_o^2}\right)\left(1-\frac{2M}{r_o}\right)},
\end{equation}
\nin and
\begin{eqnarray}
\cl_o^2 =&&\frac{2Mr_o^2}{r_o-3M}+\Bigg[\frac{qr_o\left(r_o-2M\right)}{\left(r_o-3M\right)^2} \non  \\
&&\left(q-\sqrt{4r_o \left(r_o-3M\right)+q^2}\right)\Bigg]. \label{lq}
\end{eqnarray}

\begin{figure}[h!]
  \centering
  \includegraphics[width=0.45\textwidth]{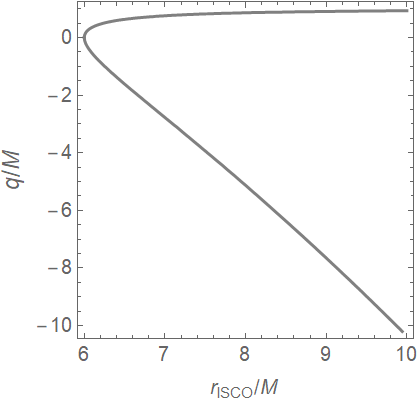}
  \caption{The radius of the ISCO vs. the charge parameter $q$.}\label{iscoq}
\end{figure}

\begin{figure}[h!]
  \centering
  \includegraphics[width=0.45\textwidth]{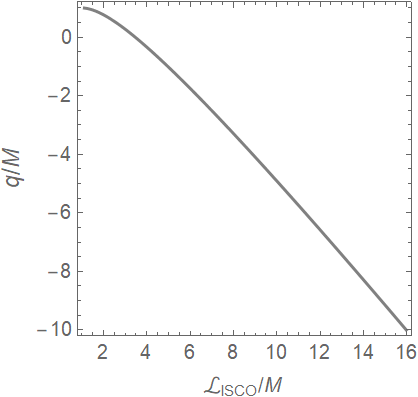}
  \caption{The specific azimuthal angular of the ISCO vs. the charge parameter $q$.}\label{Liscoq}
\end{figure}

\begin{figure}[h!]
  \centering
  \includegraphics[width=0.45\textwidth]{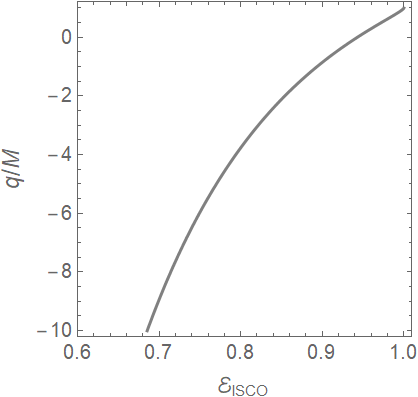}
  \caption{The specific energy of the ISCO vs. the charge parameter $q$.}\label{Eiscoq}
\end{figure}

\nin The condition for a circular orbit to be an ISCO $V''(r_o)=0$ reads
\begin{eqnarray}
&&[r_o \left(\cl_o^2+r_o^2\right)(r_o-2M)]^{1/2}\left(\cl_o^2-2 Mr_o\right)+\;\;\;\;\;\;\;\non \\
&&q\left[\cl_o^2(r_o-M)+r^2(2r_o-3M)\right]=0.
\end{eqnarray}
\nin Figs.~\ref{iscoq},~\ref{Liscoq} and~\ref{Eiscoq} show how the radius, azimuthal specific angular momentum and specific energy of the ISCO vary with $q$, respectively. For $q>0$, $r_\text{ISCO}$ increases very steeply and approaches infinity as $q$ approaches $M$, in agreement with Refs.~\cite{Zaj1,PQR}. This is possibly due to that the Coulomb repulsion makes the circular orbits less stable. When $q<0$, $r_\text{ISCO}$ also increases as the magnitude of $q$ increases, but more smoothly. Overall, the black hole charge pushes away $r_\text{ISCO}$ beyond $r=6M$ for both signs of $q$.

\nin At $q=M$, $\cl_\text{ISCO}$ has its minimum of $M$. It then increases monotonically as $q$ decreases. On the other hand, $\ce_\text{ISCO}$ has its maximum of $1$ when $q=M$. As $q$ decreases, $\ce_\text{ISCO}$ monotonically decreases and approaches $0$ as $q$ approaches $-\infty$. Therefore, the efficiency of energy liberation of a charged particle at the ISCO can be close to 100\% of the particle's rest energy.      



\nin In the case when $q<<-M$, we can write approximate expressions for $r_{\text{ISCO}}$, $\cl_{\text{ISCO}}$ and $\ce_{\text{ISCO}}$ as     
\begin{eqnarray}
 r_{\text{ISCO}}&\approx&\sqrt[3]{2Mq^2}, \\
\cl_{\text{ISCO}}&\approx&-q, \\
\ce_{\text{ISCO}}&\approx&3\sqrt[3]{-{M}/{4q}}.
\end{eqnarray}

\nin The results of this section, and the relationship between $r_{\text{ISCO}}$ and $q$ in particular, demonstrates that even a trace charge on a black hole can have profound astrophysical implications.

\section{Circular Orbits around a Magnetized Schwarzschild Black Hole}\label{s5}

\begin{figure}[h!]
  \centering
  \includegraphics[width=0.45\textwidth]{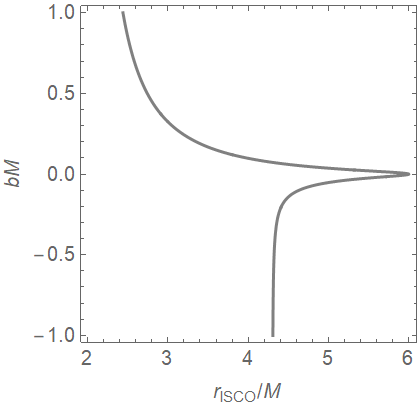}
  \caption{The radius of the ISCO vs. the magnetic parameter $b$.}\label{iscob}
\end{figure}

\begin{figure}[h!]
  \centering
  \includegraphics[width=0.45\textwidth]{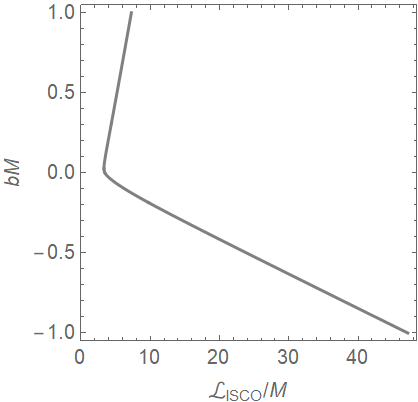}
  \caption{The specific azimuthal angular of the ISCO vs. the magnetic parameter $b$.}\label{Liscob}
\end{figure}

\begin{figure}[h!]
  \centering
  \includegraphics[width=0.45\textwidth]{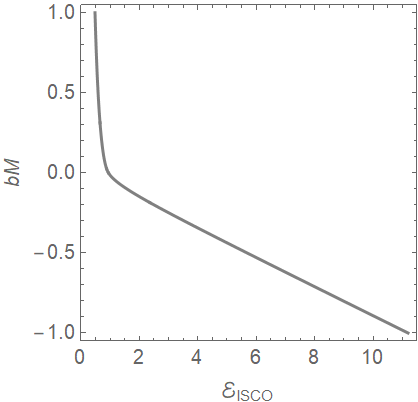}
  \caption{The specific energy of the ISCO vs. the magnetic parameter $b$.}\label{Eiscob}
\end{figure}

\nin In this section, we review the circular orbits and the ISCOs of a charged particle near a magnetized Schwarzschild black hole. Eq.~\ref{pot} with $q=0$ reads
\begin{equation}
V_+(r) = \sqrt{\left[1+\left(\frac{\cl}{r}-br\right)^2\right]f(r)}.
\end{equation}
The two conditions for circular orbits give us that
\begin{eqnarray}
\ce_o &=& \sqrt{\left[1+\left(\frac{\cl_o}{r_o}-br_o\right)^2\right]\left(1-\frac{2M}{r_o}\right)}, \\
\cl_o&=&\frac{r_o\sqrt{b^2r_o^2(r_o-2M)^2+M(r_o-3M)}}{r_o-3M}\;\;\;\;\;\;\;\; \non \\
&&-\frac{bMr_o^2}{r_o-3M}.
\end{eqnarray}
\nin For a circular orbit to be an ISCO, it must satisfy the condition
\begin{equation}
b^2r_o^3(5r_o-4M)-4b\cl_oMr-\cl_o^2+2Mr_o=0.
\end{equation} 
Fig.~\ref{iscob} shows $r_\text{ISCO}$ vs $b$. The effect of the magnetic field is always to bring the ISCO inward closer than $r=6M$. As $b$ approaches $\infty$ and $-\infty$, $r_\text{ISCO}$ approaches $2M$ and $(\sqrt{13}+5)M/2$, respectively.
Figs.~\ref{Liscob} and~\ref{Eiscob} show $\cl_\text{ISCO}$ and $\ce_\text{ISCO}$ vs. $b$ respectively. When $b\rightarrow \infty$, $\cl_\text{ISCO}$ approached $\infty$ but $\ce_\text{ISCO}$ approach $0$. The efficiency of energy release for a charged particle ending at the ISCO can be close to $100\%$ of the particle's rest energy. As $b\rightarrow -\infty$, both $\cl_\text{ISCO}$ and $\ce_\text{ISCO}$ approach $\infty$.

\section{Circular Orbits around a Charged and Magnetized Schwarzschild Black Hole}\label{s6}

\begin{figure*}[ht]
    \begin{center}
    \ba
    &&\hspace{.1cm}\includegraphics[width=0.32\textwidth]{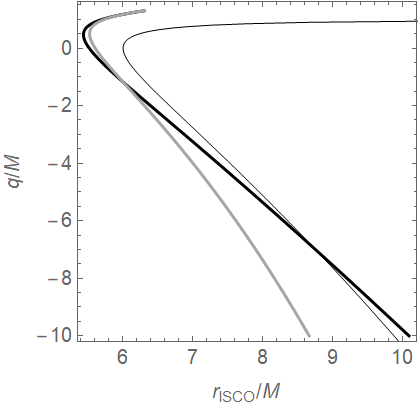}
    \hspace{.1cm}\includegraphics[width=0.32\textwidth]{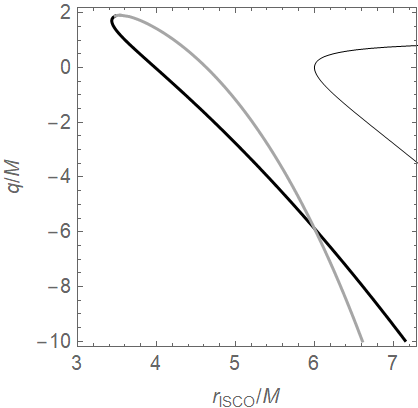}
    \hspace{.1cm}\includegraphics[width=0.32\textwidth]{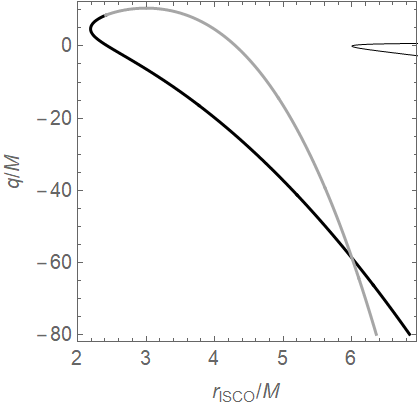}\non\\
    &&\hspace{2. cm}({\bf a })\;b=\pm 0.02/M. \hspace{2.8cm}({\bf b})\;b=\pm 0.1/M.\hspace{2.8cm} ({\bf c})\;b=\pm 1/M.\non
    \ea
    \caption{The radius of the ISCO vs. $q$ for selected values of $b$. The thick black (grey) curve corresponds to the positive (negative) value of $b$. The thin curve corresponds to $b=0$.}\label{iscoqb}
    \end{center}
\end{figure*}

\begin{figure}[h!]
  \centering
  \includegraphics[width=0.45\textwidth]{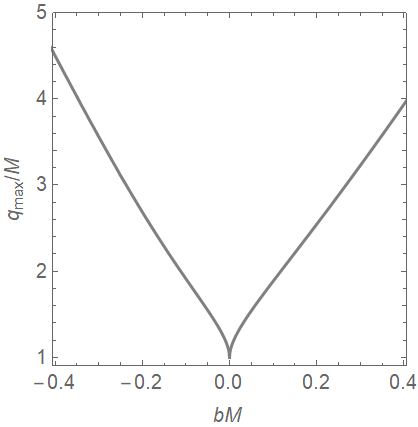}
  \caption{The dependence of the maximum value of $q$ at which the ISCO exists ($q_\text{max}$) on $b$.}\label{qmaxb}
\end{figure}

Applying the two conditions for circular orbits to the effective potential of Eq.~\ref{pot} yields

\begin{equation}
\ce_o = \frac{q}{r_o}+\sqrt{\left[1+\left(\frac{\cl_o}{r_o}-br_o\right)^2\right]\left(1-\frac{2M}{r_o}\right)},\;\;\;\;\;\;\;
\end{equation}
where $\cl_o$ is determined by the equation
\begin{eqnarray}
b^2&&r_o^4(r_o-M)+Mr_o^2(1-2b\cl_o)+\cl_o^2(3M-r_o)\;\;\;\;\; \non\\
&&-q\sqrt{r_o(r_o-2M)\left[\left(\cl_o-br_o^2\right)^2+r_o^2\right]}=0. \;\;\;
\end{eqnarray}
While it is possible to solve this equation for $\cl_o$ explicitly, the resulting expression is extremely cumbersome. The ISCO condition reads
\begin{eqnarray}
&&b^2r_o^3(5r_o-4M)-4b\cl_oMr_o-\cl_o^2+2Mr_o-\frac{q}{r_o}A \non \\
&&-\frac{q}{A}\Big[b^2r_o^4(2r_o-3M)+2b\cl_or_o^2(M-r_o)+\cl_o^2M \non \\
&&\;\;\;\;\;\;\;\:\:+r_o^2(r_o-M)\Big]=0,
\end{eqnarray} 
where
\begin{equation}
A=\sqrt{r_o(r_o-2M)\left[\left(\cl_o-br_o^2\right)^2+r_o^2\right]}.
\end{equation}
In order to visualize the behavior of $r_\text{ISCO}$ when both $q$ and $b$ are nonzero, we will reproduce Fig.~\ref{iscoq} for selected, representative values of $b$. Fig.~\ref{iscoqb} shows the effect of turning on the magnetic field on the $r_\text{ISCO}$ vs. $q$ curve. 
The magnetic field has mainly three effects on $r_\text{ISCO}$: (1) It brings $r_\text{ISCO}$ closer to the black hole. In all cases where $b\neq 0$, $r_\text{ISCO}$ is finite. (2) It increases the maximum possible value of $q$ (call it $q_{\text{max}}$) for which $r_\text{ISCO}$ exists beyond $q_{\text{max}}=M$. (3) It can creates two concurrent ISCOs when $b<0$ (see below). These three effects become more evident as $|b|$ increases. Fig.~\ref{qmaxb} shows $q_{max}$ vs. $b$. 
When $q=q_{max}$, $\cl_\text{ISCO}=0$ for $b>0$ only. The relationship between $q_{max}$ and $b$ can be well-approximated to be linear when $|b|$ $\gtrsim 10^{-1}/M$ as 
\begin{eqnarray}
q_{max}&\approx &  8.22\:bM^2 \hspace{10mm}(b\gtrsim 10^{-1}/M), \\
q_{max}&\approx & -10.4\:bM^2 \hspace{7.5mm} (b\lesssim -10^{-1}/M).
\end{eqnarray}

\nin When both $q$ and $b$ are very large and positive ($q>>M$, $b>>1/M$), $r_\text{ISCO}$ approaches $[(3+\sqrt{3})/2] M \approx 2.366\:M$ and $q_{max}$ approaches $(\sqrt{135+78 \sqrt{3}}/2)bM^2 \approx 8.217\: bM^2$. However, when $q$ and $|b|$ are very large, $q$ is positive, but $b$ is negative ($q>>M$, $b<<-1/M$), $r_\text{ISCO}$ approaches $3M$ and $q_{max}$ approaches $-6\sqrt{3}bM^2\approx -10.39\: bM^2$.
 
\nin We can see in the three plots of Fig.~\ref{iscoqb} that the positive $b$ and negative $b$ curves always cross at $r_{\text{ISCO}}=6M$, the neutral particle value. This finding was noted in Ref.~\cite{HH}. It maybe tempting to think that the radial component of the electromagnetic force ($eF^r_{\;\;\rho}u^{\rho}$) vanishes where the crossing occurs, but this is not the case. The corresponding values of $\cl_{\text{ISCO}}$ and $\ce_{\text{ISCO}}$ are different from the neutral particle values. 

\begin{figure}[h!]
  \centering
  \includegraphics[width=0.45\textwidth]{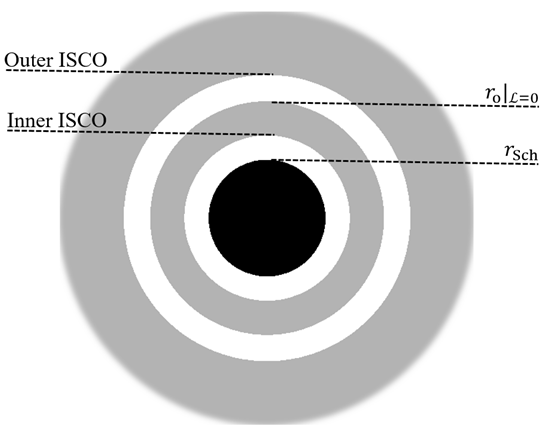}
  \caption{The inner and outer bands of stable circular orbits.}\label{disk}
\end{figure}

The most interesting finding is the occurrence of an inner and outer ISCOs when $b<0$. This happens when the charge parameter is in the interval $q_{\text{max}}(|b|)<q<q_{\text{max}}(-|b|)$. The coexistence of the two ISCOs implies the existence of a forbidden zone where stable circular orbits of charged particles cannot exist. The outer boundary of the forbidden zone is the outer ISCO. The inner boundary is the circular orbit at $r_o|_{\cl_o=0}$, the value of $r_o$ at which $\cl_o=0$. The true ISCO is the inner ISCO. Fig.~\ref{disk} demonstrate this structure. 
 
\begin{figure}[h!]
  \centering
  \includegraphics[width=0.45\textwidth]{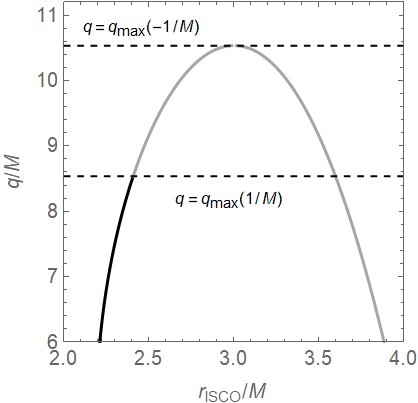}
  \caption{The radius of the ISCO vs. $q$ for $b=1/M$ (black) and $b=-1/M$ (grey). The grey curve is double-valued in the interval $q_{\text{max}}(1/M)<q<q_{\text{max}}(-1/M)$ or $8.539M<q<10.53M$.}\label{2iscos}
\end{figure}

As an illustration, let us consider the case of Fig.~\ref{iscoqb} (c), where $b = -1/M$. Fig~\ref{2iscos} is a magnification of the region where two ISCOs exist in figure Fig.~\ref{iscoqb} (c). 

\begin{figure}[h!]
  \centering
  \includegraphics[width=0.45\textwidth]{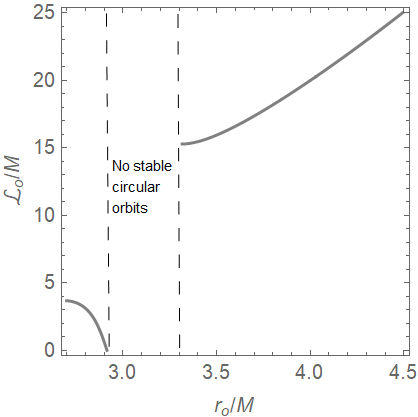}
  \caption{The radius of the stable circular orbit $r_o$ versus $\cl_o$ when $b=-1/M$ and $q=10M$. No stable circular orbits exist between the dashed lines.}\label{Lovsro}
\end{figure}

\nin Fig.~\ref{Lovsro} shows how the radius of the circular orbit $r_o$ changes with $\cl_o$ when $b=-1/M$ and $q=10M$. The inner (true) ISCO is at $r_o=2.682M$. Between $r_o=2.914M$ (the value when $\cl_o=0$) and $r_o=3.314M$ (the larger $r_{\text{ISCO}}$), no stable circular orbits exist.
 
\begin{figure*}[ht]
    \begin{center}
    \ba
    &&\hspace{.1cm}\includegraphics[width=0.32\textwidth]{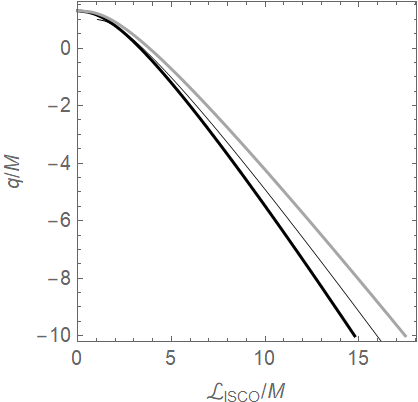}
    \hspace{.1cm}\includegraphics[width=0.32\textwidth]{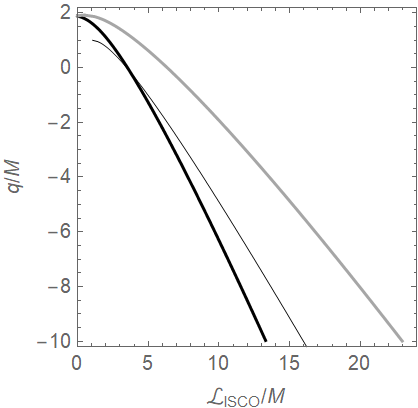}
    \hspace{.1cm}\includegraphics[width=0.32\textwidth]{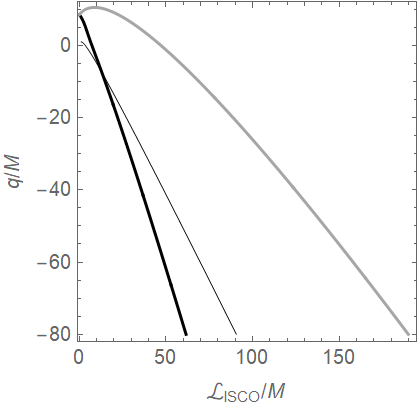}\non\\
     &&\hspace{2. cm}({\bf a })\;b=\pm 0.02/M. \hspace{2.8cm}({\bf b})\;b=\pm 0.1/M.\hspace{2.8cm} ({\bf c})\;b=\pm 1/M.\non
    \ea
    \caption{The specific azimuthal angular momentum of a charged particle at the ISCO vs. $q$ for selected values of $b$. The thick black (grey) curve corresponds to the positive (negative) value of $b$. The thin curve corresponds to $b=0$.}\label{Liscoqb}
    \end{center}
\end{figure*}
\begin{figure*}[ht]
    \begin{center}
    \ba
    &&\hspace{.1cm}\includegraphics[width=0.32\textwidth]{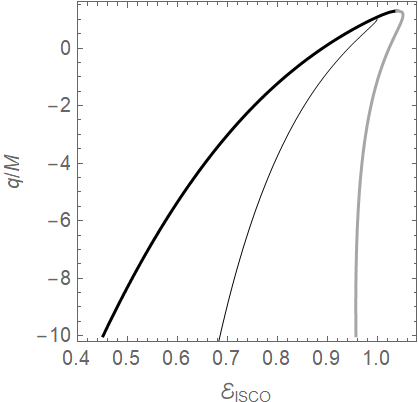}
    \hspace{.1cm}\includegraphics[width=0.32\textwidth]{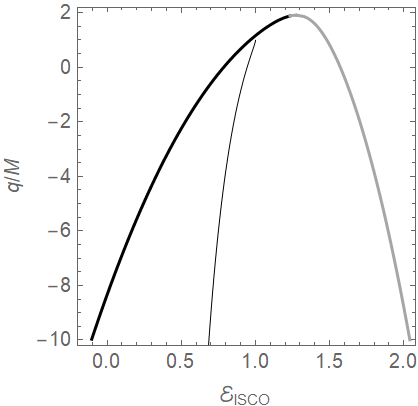}
    \hspace{.1cm}\includegraphics[width=0.32\textwidth]{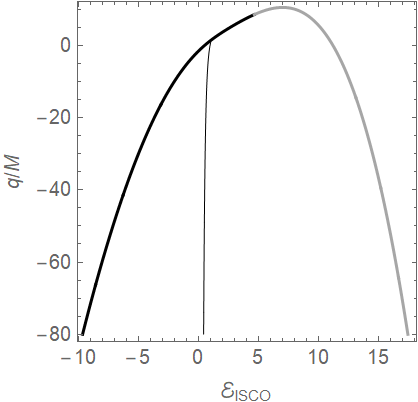}\non\\
     &&\hspace{2. cm}({\bf a })\;b=\pm 0.02/M. \hspace{2.8cm}({\bf b})\;b=\pm 0.1/M.\hspace{2.8cm} ({\bf c})\;b=\pm 1/M.\non
    \ea
    \caption{The specific energy of a charged particle at the ISCO vs. $q$ for selected values of $b$. The thick black (grey) curve corresponds to the positive (negative) value of $b$. The thin curve corresponds to $b=0$.}\label{Eiscoqb}
    \end{center}
\end{figure*}

\begin{figure}[h!]
  \centering
  \includegraphics[width=0.45\textwidth]{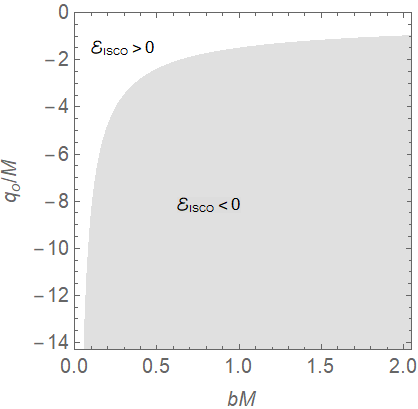}
  \caption{The critical charge parameter $q_o$ at which $\ce_{\text{ISCO}}$ vanishes vs. $b$. For $q<q_o$ (grey region), $\ce_{\text{ISCO}}$ is negative.}\label{qovsb}
\end{figure}

\nin Figs.~\ref{Liscoqb} and~\ref{Eiscoqb} show $\cl_\text{ISCO}$ and $\ce_\text{ISCO}$ that correspond to $r_\text{ISCO}$ shown in Fig.~\ref{iscoqb}. It is compelling that $\ce_\text{ISCO}$ becomes negative when $b>0$ after $q$ falls behind a certain negative value, call it $q_o$. Fig.~\ref{qovsb} shows how $q_o$ changes with $b$. The parameter $q_o$ approaches $0$ ($-\infty$) as $b$ approaches $\infty$ ($0$). We can write approximate expressions for $q_o$ when $b<<1/M$ or $b>>1/M$. They read
\begin{eqnarray}
q_{o} &\approx &  -\sqrt{\frac{8\sqrt{3}M}{9b}}=-1.241\sqrt{\frac{M}{b}} \hspace{1mm} (b>>1/M),\;\;\;\;\;\;\; \\
q_{o} &\approx &  -\frac{1}{2b}		\hspace{35mm} (b<<1/M).
\end{eqnarray}

\nin This finding may have important astrophysical consequences. The energy liberated by a charged particle ending in such an ISCO can be several order of magnitudes greater than its rest energy! Similar 'superbound' stable circular orbits were found near magnetized Kerr black holes in Ref.~\cite{Z}. However, we are not aware of any negative-energy orbits outside the event horizon of the Schwarzschild black hole.  

\section{Summary}\label{sum}

We have studied the ISCOs of charged particles near a weakly charged and magnetized Schwarzschild black hole. The effect of the black hole's charge alone is to push the ISCO beyond the neutral particle's ISCO, regardless of whether the Coulomb force is repulsive or attractive. When the Coulomb force is repulsive, the ISCO does not exist beyond some critical ratio of the Coulomb force to the 'gravitational force'. 
The binding energy of a charged particle at the ISCO can be as much as the particle's rest energy.

The effect of the magnetic field alone is to bring the ISCO closer than the neutral particle ISCO in all cases. When the magnetic force is radially out, the particle's ISCO can approach the event horizon where the particle's binding energy approaches its rest energy. 

The problem becomes much richer when the black hole is both charged and magnetized. The charge and magnetic field have competitive effects on the ISCO's radius. The critical ratio of the Coulomb force to the 'gravitational force' beyond which the ISCO does not exit becomes greater as the magnetic field becomes stronger. An interesting result is that the particle's energy can be negative. The energy liberation in such cases can be several orders magnitude greater than particle's rest energy. 

The most interesting result is the possibility of the existence of two bands of charged particles' circular orbits, separated by a region of no stable circular orbits.

The problem can be more sophisticated and more astrophysically interesting when restudied in Kerr spacetime. The black hole rotation can have significant effects on the energy libation efficiency, position of the ISCOs and hence the forbidden zone. Another important modification is to consider more realistic magnetic fields. 

\section*{Acknowledgment}

The author gratefully acknowledges the deanship of academic research at King Fahd University of Petroleum and Minerals for financially supporting this work under project code SR181025.

\end{document}